# Synthesis and physical characterization of superconductivity-magnetism crossover compound $RuSr_2EuCeCu_2O_{10-\delta}$

V.P.S. Awana[1,2,*], S. Balamurugan[2], L. S. Sharath Chandra[3], Akshay Deshpande[1], V. Ganesan[3], H. Kishan[1], E. Takayama-Muromachi[2] and A.V. Narlikar[3]

[1]National Physical Laboratory, Dr. K.S. Krishnan Marg, New Delhi-110012, India
[2]Superconducting Materials Center, NIMS, 1-1 Namiki, Tsukuba, Ibaraki, 305-0044, Japan
[3]UGC-DAE Consortium for Scientific Research, University Campus, Khandwa Road, Indore-452017, MP, India

We carefully studied the nonsuperconducting sample of the magneto-superconducting $RuSr_2(Eu_{1-x}Ce_x)Cu_2O_{10-\delta}$ series with composition $RuSr_2EuCeCu_2O_{10-\delta}$. This compound seems to exhibit a complex magnetic state as revealed by host of techniques like resistivity, thermopower, magnetic susceptibility and *MR* measurements. The studied compound exhibited ferromagnetic like *M(H)* loops at 5, 20 and 50 K, and semiconductor like electrical conduction down to 5 K, with – $MR^{7Tesla}$ of up to 4% at low temperatures. The $–MR^{7Tesla}$ decreases fast above 150 K and monotonically becomes close to zero above say 230 K. Below, 150 K $–MR^{7Tesla}$ decreases to around 3% monotonically down to 75 K, with further increase to 4% at around 30K and lastly having a slight decrease below this temperature. The thermopower *S(T)* behavior closely followed the $–MR^{7Tesla}$ steps in terms of d(*S/T*)/d*T* slopes. Further, both $MR^{7Tesla}$ steps and d(*S/T*)/d*T* slopes are found in close vicinity to various magnetic ordering temperatures ($T_{mag.}$) of this compound.

PACS No. 74.72. Jt, 74.25. Bt, 75.80. +q

Key words: Ru-1222 magneto-superconductors, Thermopower, Magnetization, and Magneto-Resistance

## INTRODUCTION

Recently, exciting observations of negative lattice expansion (*NLE*) were reported for nonsuperconducting ruthenocuprate $RuSr_2Y_{0.2}Nd_{0.9}Ce_{0.9}Cu_2O_{10}$ by Mclaughlin et al. [1], which were however retracted [2] due to erroneous analysis of the neutron diffraction data that they subsequently discovered. Magneto-superconductivity of rutheno-cuprates viz. $RuSr_2(Ln,Ce)Cu_2O_{10-\delta}$ (Ru-1222) [3] and $RuSr_2LnCu_2O_{8-\delta}$ (Ru-1212) [4] is currently an area of intense research and also a rich field for topical reviews [5-7]. The mainstay of the current research seems to be centered around the magnetic state of these compounds and the dramatic changes in magnetic behavior that take place on cooling below the room temperature. Consequently, much efforts are being directed towards gaining an insight into the un-doped and magnetic ground state of these compounds [8,9]. Perhaps a possible path breaker in the field has been the recent report of Mclaughlin et al. [1] who, for the nonsuperconducting ferromagnetic compound $RuSr_2Y_{0.2}Nd_{0.9}Ce_{0.9}Cu_2O_{10}$, of Ru-1222 series, doped near the edge of the boundary between antiferromagnetism and superconductivity, observed an unusual negative lattice expansion (*NLE*) below the magnetic ordering temperature of Ru and a high magneto-resistance value exceeding -30% at 5 K.

Though the (*NLE*) being reported for nonsuperconducting ruthenocuprate $RuSr_2Y_{0.2}Nd_{0.9}Ce_{0.9}Cu_2O_{10}$ by Mclaughlin et al. [1] was retracted [2], the same stimulated renewed interest in physical characterization of superconductivity-magnetism crossover ruthenocuprates viz. $RuSr_2EuCeCu_2O_{10-\delta}$. It is worth noting that $RuSr_2EuCeCu_2O_{10-\delta}$ is the parent compound of the magneto-superconducting Ru-1222 family. It is worth pointing out that the compound studied in ref. 1, i.e. $RuSr_2Y_{0.2}Nd_{0.9}Ce_{0.9}Cu_2O_{10}$ was under-doped and magnetic that did not turn superconducting at lower temperature, and in this respect our presently studied material $RuSr_2EuCeCu_2O_{10-\delta}$ is essentially similar but with an important difference that it is comparatively more optimally doped to remain even much closer to the boundary separating the magnetic and superconducting states. This will be discussed later when we compare the resistivity and the negative *MR* of these two compounds. In reality not only the Ce content but overall oxygen of these compounds decide the carrier density in them [10]. Higher resistivity and negative *MR* of up to -25% had already been reported by some of us for $N_2$-annealed $RuSr_2Gd_{1.6}Ce_{0.4}Cu_2O_{10-\delta}$ [10]. We present the electrical, magnetic and thermal characterization of the compound. Our findings, we believe, are directly relevant for understanding the mysterious magnetism of rutheno-cuprates.

## EXPERIMENTAL DETAILS

The $RuSr_2EuCeCu_2O_{10-\delta}$ sample is synthesized through a solid-state reaction route from $RuO_2$, $SrO_2$, $Eu_2O_3$, $CeO_2$ and CuO. Calcinations were carried out on the mixed powder at 1000, 1020, 1050 and 1080

$^0$C each for 24 hours with intermediate grindings. The pressed bar-shaped pellets were annealed in a flow of oxygen at 1085 $^0$C for 40 hours and subsequently cooled slowly over a span of another 20 hours down to room temperature. Powder x-ray diffraction (XRD) data was collected using a powder diffractometer with CuK$_\alpha$ radiation. Magnetization measurements were carried out on a superconducting-quantum-interference-device (SQUID) magnetometer (Quantum Design: MPMS-5S). Resistivity measurements were made under magnetic field in the temperature range of 5 to 300 on Quantum Design PPMS (Physical Property Measurement System). Thermoelectric power (TEP) measurements were carried out by dc differential technique over a temperature range of 5 – 300 K, using a home made set up. Temperature gradient of ~1 K was maintained throughout the TEP measurements.

**RESULTS AND DISCUSSION**

Fig. 1 shows the powder diffraction patterns of RuSr$_2$EuCeCu$_2$O$_{10-\delta}$ at 293 K. Felner et al [11] reported similar results on RE$_{1.5}$Ce$_{0.5}$RuSr$_2$Cu$_2$O$_{10}$ (RE = Eu, Gd). The powder diffraction pattern resembles that of other iso-structural compounds reported in the literature [12,13]. Barring some very weak reflections, the entire pattern can be indexed on the basis of a body centered tetragonal cell with space group *I*4/*mmm*. The lattice parameters are a =3.8402(2)Å  c= 28.516(2)Å, which are in good agreement with earlier reports [11].

DC magnetic susceptibility versus temperature χ(*T*) plot for RuSr$_2$EuCeCu$_2$O$_{10-\delta}$ compound is shown in figure 2, in both zero-filed-cooled (ZFC) and field-cooled (FC) situations. The applied field was 5 Oe. It can be seen from the figure that χ$^{ZFC}$ and FC is almost constant till 160K, below which both exhibit a sharp rise. However, on further lowering of temperature χ$^{ZFC}$ takes a down turn peak-like-shape at around 120K, the χ$^{FC}$, on the otherhand, goes on increasing, though at a slower rate below 120K, before saturating below say 30 K. Both χ$^{ZFC}$ and χ$^{FC}$ remain positive over the complete temperature range down to 5 K. We kept a low value of 5 Oe for the applied field for these measurements since many of the interesting magnetic features tend to readily disappear at higher applied fields. This was apparently the case with RuSr$_2$Y$_{0.2}$Nd$_{0.9}$Ce$_{0.9}$Cu$_2$O$_{10}$, which was studied at a relatively much larger field 0.10 Tesla [1]. The temperature variation of χ$^{ZFC}$ and χ$^{FC}$, reported above, is similar to the other Ru-1222 magneto-superconducting samples of the present series [3,8,10,13]. The only difference in the present case is that diamagnetic signal is not seen in χ$^{ZFC}$ at low temperatures, because the sample studied is not superconducting. Some of these features are described and corroborated separately [15]. Basically the Ru spins order anti-ferro-magnetically (AFM) at around 180 K, develop a ferromagnetic component within at around 150 K and a spin glass like (SG) structure at χ$^{ZFC}$ cusp below 120 K. The FM like *M(H)* loops of the compound are shown in inset of figure 2 at various temperatures and in applied fields of up to 3000 Oe. It is evident from the inset figure that the remanant magnetization $r_m$ and the coereceive field ($H_c$) are around 0.8 $\mu_B$ and 300 Oe respectively at 5 K. Both $r_m$ and $H_c$ are not much changed at 20 K and 50 K. However at a higher temperature of 100 K, though the $r_m$ is visible with very low value, the $H_c$ is not seen, as the *M(H)* loop is practically closed. At 150 K and above, the compound exhibits more or less a paramagnetic like *M(H)* behavior. However it is not exactly paramagnetic as the AFM correlations are present above 150 K, as evidenced by other techniques [16]. When compared directly with magneto-superconducting composition of the same series i.e. $RuSr_2Eu/Gd_{1.5}Ce_{0.5}Cu_2O_{10-\delta}$ compounds [17,18], the magnetic ordering temperatures in terms of $T_{AFM}$, $T_{SG}$ and $T_{FM}$ are found to be higher by around 20-30 K for the currently studied ground state nonsuperconducting $RuSr_2EuCeCu_2O_{10-\delta}$ compound. This is in agreement with an earlier report of ours on highly under-doped $N_2$-annealed nonsuperconducting sample of composition $RuSr_2Gd_{1.6}Ce_{0.4}Cu_2O_{10-\delta}$, which exhibited higher $T_{mag}$ in comparison to the superconducting rutheno-cuprates.

Fig. 3 depicts the resistivity versus temperature $\rho$ (*T*) measurements for the studied $RuSr_2EuCeCu_2O_{10-\delta}$ compound. Room temperature resistivity $\rho_{300K}$ is found to be 29.08 m$\Omega$.cm. To estimate the extent of under-doping in the compound we compare the same with the reported data [1] on $RuSr_2Y_{0.2}Nd_{0.9}Ce_{0.9}Cu_2O_{10}$ for which the corresponding value is 50-100 m$\Omega$.cm. This shows that extent of under-doping is more on $RuSr_2Y_{0.2}Nd_{0.9}Ce_{0.9}Cu_2O_{10}$ than in currently studied $RuSr_2EuCeCu_2O_{10-\delta}$ compounds and thus the former seems comparatively farther away from the border of the magnetism-superconductivity crossover, than the latter. It therefore seems that our sample is just at the close proximity of becoming magneto-superconducting. Incidentally, for superconducting $RuSr_2Eu/Gd_{1.5}Ce_{0.5}Cu_2O_{10-\delta}$ compounds $\rho_{300K}$ is around 10 m$\Omega$.cm [17, 18]. The $\rho$ (*T*) behavior of the compound is found to be semiconductor like, as expected for any other under-doped rutheno-cuprate system.

The insets of Fig. 3 show the *MR* at 5 K in various fields in left side and the same in right side at fixed field (7 Tesla) and varying temperature from 300 down to 5 K. At 5 K the $-MR^{7Tesla}$ of around 4% is seen. Earlier in a highly under-doped $N_2$-annealed sample of composition $RuSr_2Gd_{1.6}Ce_{0.4}Cu_2O_{10-\delta}$, we reported $-MR^{7Tesla}$ of above 25% at 5 K, which increased further at 2 K [10]. Interestingly this is very close to the $-MR^{7Tesla}$ of above 30% for $RuSr_2Y_{0.2}Nd_{0.9}Ce_{0.9}Cu_2O_{10}$ sample of ref. [1]. This again indicates that the presently studied sample is truly the one to be close to superconductivity – magnetism crossover point. The fixed field (7 Tesla) and varying temperature (300 K – 5 K) *MR* shown on the right side inset of the figure, exhibit an interesting behavior in terms of approaching a maximum value of -4% at around 150

K. The $-MR^{7Tesla}$ decreases fast above 150 K and monotonically becomes close to zero above say 230 K. Below, 150 K $-MR^{7Tesla}$ decreases to around 3% monotonically down to 75 K, with further increase to 4% at around 30K and lastly having a slight decrease below this temperature. The temperature 75 K lies between $T_{SG}$ and $T_{FM}$, and the $-MR^{7Tesla}$ turning point at this temperature is mostly due to the presence of competing magnetic phase separated SG and canted FM regions. This behavior of $-MR^{7Tesla}$ is very different from that observed for either $N_2$-annealed Ru-1222 (10) or $RuSr_2Y_{0.2}Nd_{0.9}Ce_{0.9}Cu_2O_{10}$ (1). For the Ru-1222 compounds studied earlier in refs. 1 and 10, the $-MR^{7Tesla}$ increases monotonically on cooling, reaching the highest value of up to 30%, without having any non-monotonic steps. Interestingly the temperatures corresponding to the non-monotonic variation of $-MR^{7Tesla}$ seen with the present sample is consistent with the various magnetic ordering temperatures of the compound. In particular $-MR^{7Tesla}$ is nearly negligible above $T_{AFM}$, hits a maximum below around $T_{SG}$ with a further decrease and later a near constant value in most of FM region. As discussed in magnetism part the magnetic structure of the compounds seems quite complex hence perhaps due to the presence of overlapping magnetic phase separated regions in terms $T_{AFM}$, $T_{SG}$ and $T_{FM}$, the $-MR^{7Tesla}$ does not follow exactly these temperatures, but the regions of the overlap. For example though the $T_{SG}$ occurs at above 100 K, the $-MR^{7Tesla}$ turning step is seen at around 75 K. Though the exact magnetic structure of these compounds as probed by neutron scattering experiments is yet warranted, some of us earlier [15] had shown clearly that the AFM ordered spins of Ru turns into a spin glass like structure before finally having the canted ferromagnetism. We believe that the nonmonotonicity of $-MR^{7Tesla}$ with temperature in the present nonsuperconducting $RuSr_2EuCeCu_2O_{10-\delta}$ compound is possibly due to the magnetic phase separations occurring at three distinct ordering temperatures in this class of compounds.

We also measured thermo-electric power (*S*) at varying temperatures for our sample. The *S(T)* plot of the compound was similar to that as for other rutheno-cuprates but without superconductivity transition. The diffusion thermopower may be obtained form Mott formula [19] as

$$S = (\pi^2 k_B^2 T / 3e\sigma(E_F))(d\sigma(E_F)/dE)$$

The temperature dependence *S/T* for the above formula may approximately characterise the temperature dependence of resistivity while derivative of S/*T* with respect to temperature yield information about the effect of change in fermi surface on thermopower [20].

The thermopower behavior of $RuSr_2EuCeCu_2O_{10}$ is similar to most of the other copper oxide ceramics with a broad hump at 240 K below which thermopower falls off gradually (Fig.3). It is known that many of the underdoped cuprate ceramics, exhibit a broad hump in thermopower, similar to the present case, at high temperatures due to the formation of pseudo gap [21]. A monotonically decreasing

TEP has also been observed in spite of semiconducting resistivity behavior in the underdoped cases of $Y_{0.6}Pr_{0.4}Ba_{2-x}Sr_xCu_3O_7$ system [22].

From the $S$ v/s $T$ plot (Fig.4) no clear signatures, corresponding to the antiferromagnetic and ferromagnetic transitions, are observed. However a plot of d$S$/d$T$ with temperature (the inset of Fig 4) clearly reveals a kink at 180 K which may correspond to AFM transition. A peak in d$S$/d$T$ is also observed at 130 K, which may be linked with the FM transition. A deviation in $S$ v/s $T$ and a corresponding clear change in slope in d$S$/d$T$ compared to other insulating cuprates is observed below 80 K probably signifying *NLE*. We have also observed drastic changes in the variation of slope at different ordering temperatures in the d($S$/$T$)/d$T$ versus $T$ plot (plot not shown) indicating changes in the Fermi surface at these ordering temperatures.

As temperature is decreased the thermopower decreases monotonically to zero even in the semiconducting state in the present case and may be due to the onset of coherent antiferromagnetic lattice. Phonon drag peak, associated with the lattice, even if present will not be evident in these samples because of the development of coherent antiferromagnetic lattice. It may be mentioned that the FM state in this compound arises from canting of the spins out of an AFM state [23], and an electronically phase separated system permits the both to coexist. It is interesting to note that Wang and Ong [24] explained a similar vanishing of the thermopower, in an under-doped YBCO with oxygen below 6.4, in terms of particle-hole symmetry in the antiferromagnetic state. The thermopower is expected to vanish if the conductivity function displays exact particle hole symmetry about chemical potential (µ). Even though the above said arguments are not exhaustive it is quite sufficient to explain the behavior seen. The upward deviation seen below 75 K develops to a small hump peaking at about 30 K before thermopower drops down to zero may be attributed to increase in entropy due to spin glass freezing. Finally a remark with a caution is about the appearance of a sharp peak in d($S$/$T$)/d$T$ below 20 K may signify the presence of a narrow density of states at fermi level [25].

The $S^{300K}$ value is around 63 µV/K, which is three times higher to that of superconducting $RuSr_2Eu_{1.5}Ce_{0.5}Cu_2O_{10-\delta}$ compound [18], which is same as the compared resistivity ratios discussed above. Discussion of $\rho_{300K}$ and $S^{300K}$ is important in assessing the crossover proximity of the sample in relation to magnetism-superconductivity interface. Also the same is supposed to have an effect on the extent of low temperature *MR*.

In conclusion the magnetism – superconductivity crossover phase pure compound $RuSr_2Eu_{1.5}Ce_{0.5}Cu_2O_{10-\delta}$ is synthesized and its magnetization, magneto resistance and thermoelectric power is studied. This compound seems to exhibit a complex magnetic state, which needs to be explored fully through neutron scattering experiments. Unfortunately the only available neutron scattering results on this compound [1] are retracted [2].

This work is partially supported by INSA-JSPS bilateral exchange visit of Dr. V. P. S. Awana to NIMS Japan

**FIGURE CAPTIONS**

Figure 1   Powder XRD patterns of $RuSr_2EuCeCu_2O_{10-\delta}$ compound.

Figure 2   $\chi(T)$ plot for $RuSr_2EuCeCu_2O_{10-\delta}$ compound, inset shows the $M(H)$ for the same

Figure 3   $\rho(T)$ plot for $RuSr_2EuCeCu_2O_{10-\delta}$ compound, inset shows the $M(R)$ for the same

Figure 4   $S(T)$ plot in main panel and in inset the $dS/dT$ versus $T$ plot for $RuSr_2EuCeCu_2O_{10-\delta}$ compound

Fig. 1

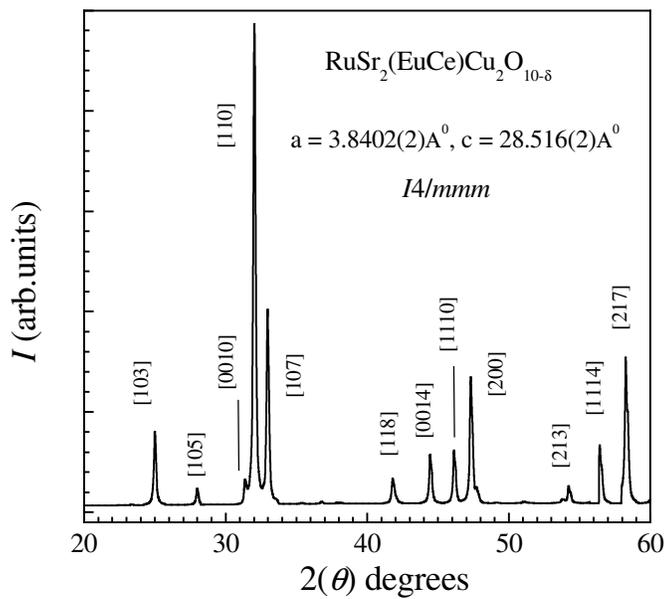

Fig. 2

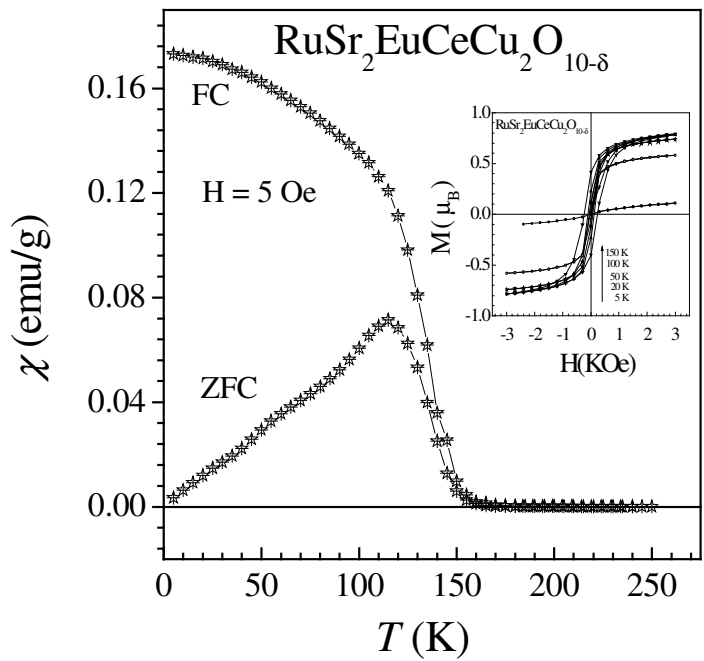

Fig. 3

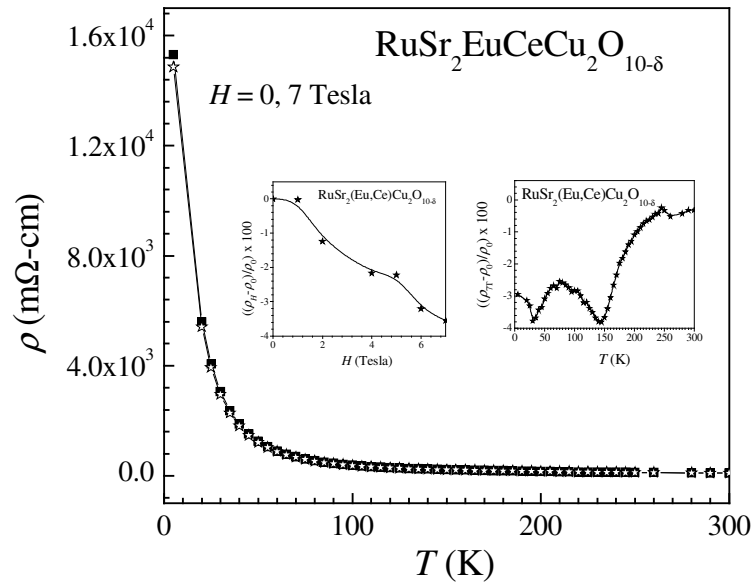

Fig. 4

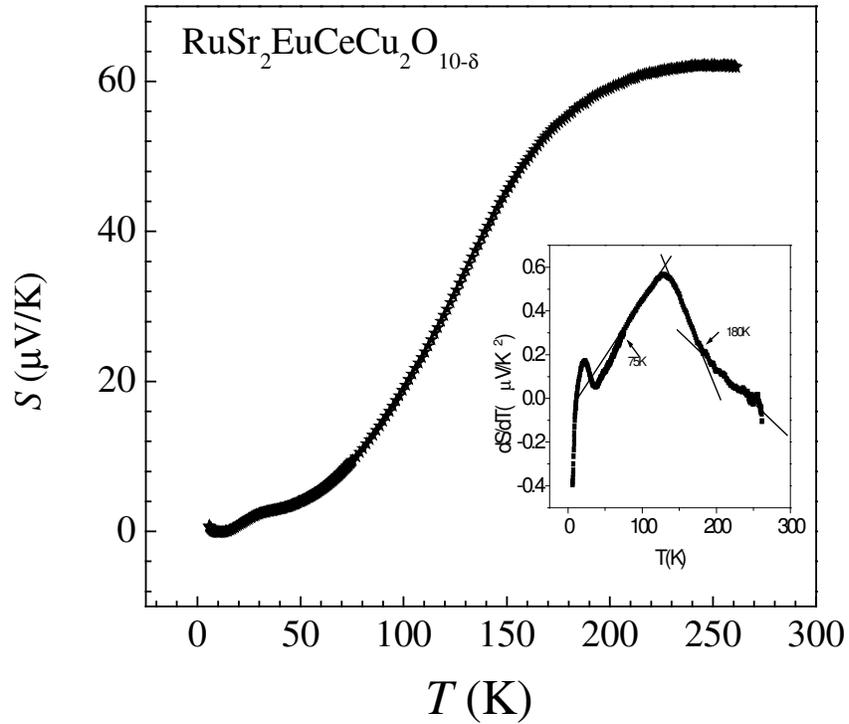

11